\documentclass[12pt,a4paper,twoside]{article}
\usepackage{epsfig}
\usepackage{axodraw}
\newlength{\figwidth}
\newlength{\figheight}
\def\z0{Z}
\def\beq{\begin{equation}}
\def\eeq{\end{equation}}
\def\beqar{\begin{eqnarray*}}
\def\eeqar{\end{eqnarray*}}
\def\earr{\end{array}}
\def\bfi{\begin{figure}}
\def\efi{\end{figure}}
\def\btab{\begin{table}}
\def\etab{\end{table}}
\def\bce{\begin{center}}
\def\ece{\end{center}}

\def\lsim{\:\raisebox{-0.5ex}{$\stackrel{\textstyle<}{\sim}$}\:}
\def\gsim{\:\raisebox{-0.5ex}{$\stackrel{\textstyle>}{\sim}$}\:}
\def\etal{{\it et al.}}
\newcommand{\ra}{\rightarrow}
\newcommand{\bi}{\bibitem}
\def\eg{{\rm e.g. }}
\def\ie{{\rm i.e. }}

\title{
       \vspace{-1.5cm}
       \begin{flushright}
       \begin{tabular}{l}
       {\large CERN-TH/2002-184 }    \\[-3mm]
       {\large IFT - 32/2002}    \\[-3mm]
       {\large hep-ph/0208076 }  \\[-3mm]
       {\large August 2002}
       \end{tabular}
       \end{flushright}
       \vspace{1.5cm}
      \sc  Precision muon $g-2$ results and light Higgs bosons in the 2HDM(II) 
              }
 \author{
 Maria Krawczyk \\
 {\small\it  Theory Division, CERN, CH-1211 Geneva 23, Switzerland}\\ [-2mm]
{\small and} \\[-2mm]
 {\small\it Institute of Theoretical Physics, Warsaw University, 
        ul. Ho\.za 69, 00-681 Warsaw, Poland} 
 } 

\date{}
\begin{document} 

\maketitle 

\vfill

\begin{abstract}
We discuss  the implications of the recent  evaluations of the SM contribution
to  ~$a_{\mu}=~(g-2)_{\mu}/2$ in the light of the latest  E821 measurement,
which indicate  $\sim$ 3$\sigma$ deviation.
We derive the 95\% CL interval, $\delta a_{\mu}$, to be used to constrain any additional contribution to $a_{\mu}$, beyond the SM ones; it  has to have   a positive sign.
We apply the $\delta a_{\mu}$ to constrain
 the light Higgs-boson scenarios in a   
Two-Higgs-Doublet-Model  (``Model II''). 
If   the constraints from the new  $(g-2)_{\mu}$ results 
are combined with  the other existing  constraints,   
one can   exclude a light-scalar scenario at  95\% CL, while a 
light-pseudoscalar
scenario can be realized, 
for a pseudoscalar mass between 25 and 70  GeV, with $\tan \beta$ 
in the range 
$25 \lsim \tan \beta \lsim 115$.\\ 
\end{abstract}

\begin{center}
{\it Dedicated to Stefan Pokorski on the  occasion of his 60th birthday

To appear in Acta Physica Polonica B}
\end{center}
\vfill

\begin{flushleft} 
CERN-TH/2002- 184 \\
IFT - 32/2002    \\
hep-ph/0208076  \\
August 2002
\end{flushleft}

\hrule
{\small \it  Supported in part by Polish Committee for Scientific Research,
 Grant  5 P03B 121 20 (2002),
and by the European Community's Human Potential
Programme under contract HPRN-CT-2000-00149 Physics at Colliders.}
\thispagestyle{empty}

\clearpage
  
\section{Introduction}
A  precision measurement of the $g-2$ for the muon at BNL is expected 
to test  the electroweak (EW) sector of the Standard Model (SM)
 and  at the same time to shed  light on  possible effects of 
 ``new physics". 
After a release of the  new E821 result \cite{BNL2002},
based on  the  $\mu^+$ data collected in the year 2000,  
a current mean of experimental results  
for $(g-2)_{\mu}$    is (\cite{BNL2002}:
\begin{equation}
a_{\mu}^{exp}\equiv{{(g-2)_{\mu}^{exp}}\over{2}}=11~659~203~(8)
\times 10^{-10},
\end{equation}
with an  uncertainty  (in parentheses) that  is  almost two times  smaller 
than  in the previous measurement \cite{Brown:2001mg},
and only two times larger than the ultimate goal of the E821 experiment.

The Standard Model prediction for $a_{\mu}$ 
consists of the QED,  EW  and hadronic contributions:
$$a_{\mu}^{SM}=a_{\mu}^{QED}+a_{\mu}^{EW}+a_{\mu}^{had}.$$
The  QED contribution, which constitutes the bulk of the SM contributions,
is calculated  up to 
four loops and the ${\cal O }(\alpha^5)$ term is estimated \cite{QED}; 
its  uncertainty is very small $\sim$ $3 ~\times ~10^{-11}$. 
The EW contribution,  based on  one- and two-loop diagrams,
is also known to a  similar accuracy [4--8]
 The  EW contribution  is small 
(152 $\times ~10^{-11}$)
 and is  only about two times larger than the present experimental uncertainty 
(see (1)).
The hadronic contribution, $\sim$ 7000 $ \times ~10 ^{-11}$, 
is the second largest  contribution to $a_{\mu}^{SM}$. 
It is known  at  present with an accuracy of 1 \%. This  uncertainty
is the  main source of the   uncertainty in  the SM prediction.
Predictions for the hadronic contribution [9-21]
 differ among themselves, 
however, these differences seem to be  much less significant than, say,
one year ago (see \eg
discussions in [11,12,17-21], and also in \cite{Czarnecki:2001pv,
Krawczyk:2001pe}).
The dominant  contribution to $a_{\mu}^{had}$, as well as one of 
 the dominant errors in its value, 
come  from the leading  vacuum-polarization (vp1) term. 
Some preliminary results of the improved calculations of  
the vp1 have been presented recently \cite{jeg02,tt}. 
These are data-driven analyses using the most recent data on hadron 
production in  $e^+e^-$ collisions
from  BES, CMD-2, SND  \cite{datalow}.
The corresponding uncertainties for the vp1, $\sim 58 ~\times ~10^{-11}$, 
are now even smaller than 
those  obtained previously by using the data on 
 $\tau $ decays  in addition to the then available
$e^+e^-$ data, \eg see  \cite{Davier:1998si,Narison:2001jt}.  
Another important issue has been the hadronic contribution   to the light-by-light (lbl) scattering, and its contribution to $(g-2)_{\mu}$. 
After  a  sign error, first pointed out  in \cite{lbl-new-first},
 was found in the earlier calculations of this contribution, a
few re-evaluations \cite{lbl-sign+} of this part have appeared during the 
last few months. All of them confirm the finding of  \cite{lbl-new-first},
\ie  a positive sign of the lbl contribution. 
The central value for the lbl varies  from 80 to $110 ~\times ~10 ^{-11} $,
depending on the analysis. Since this  contribution  can be estimated
only on a purely theoretical ground, 
it has  a sizeable uncertainty  
of the  order of  $40  \times 10 ^{-11} $ \cite{nyffeler} (or maybe even 
larger, as discussed in \cite{wise}).

The latest  SM predictions and    the present  world average 
of the experimental result (1)
differ by  $\sim$ 3$\sigma$ (a  
theoretical and experimental error combined in quadrature), 
if   preliminary results of the evaluation
of  the vp1 contribution from \cite{jeg02,tt} are used
together with the estimation for the lbl given in \cite{nyffeler}.
The significant progress in the reduction of the error on the experimental 
side and the stabilization of the SM prediction of $a_{\mu}^{had}$
makes it essential for  the possible implications
of the $(g-2)_{\mu}$ results to be reanalysed. 
The issue of whether this is a signal of supersymmetry is already being
addressed \cite{nath02}. Here we focus on 
the implication  of the latest  $(g-2)_{\mu}$ results
for  the  light-Higgs boson scenarios in the
non-supersymmetric, CP-conserving  
Two-Higgs-Doublet-Model (2HDM) in  its version called ``Model II'';
this is a continuation of  earlier  studies
\cite{Krawczyk:2001pe,mk-g2,mk-lep}.

In sec. 2  we collect some results based on the recent
calculations of  $a_{\mu}^{SM}$. As a reference for the vp1 contribution 
we take the $e^+e^-$ data-driven  analysis  done by  
Jegerlehner \cite{jeg02}, FJ02,
where the new CMD-2 results were used. 
The difference between the experimental data and the SM  
prediction, $\Delta a_{\mu}=a_{\mu}^{exp}-a_{\mu}^{SM}$, 
can be used to derive  stringent  constraints on the
parameters of models, which give additional contribution(s) to $a_{\mu}$.
We  calculate  the interval $\delta a_{\mu}$, 
which can then be  used to constrain any such  contribution, at  95\% CL.
In sec. 3 we introduce the 2HDM(II), which we wish to constrain using this
interval, and discuss briefly the 
existing constraints.
In sec. 4  we  derive the 95\% CL limits on   the parameters of this model,
using  $\delta a_{\mu}$.
Section 5 contains the combined constraints, while
the  conclusions and outlook are given in sec. 6.

\section{The $g-2$ for the muon: the new experimental
\protect\\ and theoretical results}
Here  we collect
the  SM contributions (and their uncertainties),
 which we take into account in our analysis. 
First we discuss the hadronic contributions (see table 1). 
We use the higher order contribution from \cite{Krause:1997rf}
and, for the lbl scattering contribution, we take 
an estimate from \cite{nyffeler}.
The leading vacuum polarization contribution is taken from a  preliminary 
result of an  analysis by  Jegerlehner, FJ02 \cite{jeg02}, 
where  the experimental input is based only on the $e^+e^-$ data, 
including the latest  ones from CMD-2 \cite{datalow}.
We sum all the  hadronic contributions, adding in quadrature 
the corresponding errors. This leads us to   the  
result for $a_{\mu}^{had}$ given in the last row of  table 1
(which we label   by the author  of the 
analysis of the vp1 contribution).
  $$
\begin{array}{lr}
{\rm TABLE ~1}&\\
\hline\hline
{\rm Hadronic~~contribution}  &~{  {}}~[{ \rm in}~ 10^{-11}]\\  
\hline
{\rm ho ~\cite{Krause:1997rf}}   &~~~-100~~~~(6)
   \\
{\rm lbl ~\cite{nyffeler}}   & ~80~~(40)   \\
{\rm vp1  ~\cite{jeg02}}    & 6889 ~~(58)    \\
\hline\hline
{\rm had ~[FJ02]}   &6869 ~~(71) \\
\hline 
\end{array}   
$$
We take the QED and EW terms  from
\cite{QED} and \cite{EW1,Knecht:2002hr}, respectively (see table 2).
We then calculate the total SM prediction presented in this table, 
 by adding the QED and  EW to  the full hadronic contributions, 
 and by adding in quadrature the corresponding errors. This leads us to
the SM prediction (we label it, as above,  by the author  of the 
analysis of the vp1 contribution):
 \beq
{\rm [FJ02]} \hspace{2.5cm} a^{SM}_{\mu} = 116~59 1 ~726.7 ~~~(70.9) 
\times ~10 ^{-11}.
\eeq

Taking the new world average  we calculate the quantity  
$\Delta  a_{\mu}$, defined as the difference between  the central 
values of the experimental  and theoretical (SM) predictions for $a_{\mu}$.
The   error for this quantity we estimate  by adding in 
quadrature the corresponding experimental and theoretical errors,
 $\sigma =\sqrt {\sigma_{exp}^2 + \sigma_{SM}^2 }$. 
Next we calculate the  regions of $\delta a_{\mu}$, allowed at 95\% CL,
assuming  Gaussian errors.   This leads to an interval symmetric  
around $\Delta a_{\mu}$,  
quoted in  the last row of table 2.
$$
\begin{array}{lr}
{\rm TABLE ~2}&\\
\hline\hline
{\rm SM ~contribution}  &~{  {}}~[{ \rm in}~ 10^{-11}]\\  
\hline
{\rm QED}   &~~~116~584~705.7 ~~~~(2.9)
   \\
{\rm had [FJ02]}   & ~6~869.0~~~(70.7)   \\
{\rm EW}    & 152.0 ~~~~(4.0)    \\
\hline
{\rm tot}   &116~59 1 ~726.7 ~~~(70.9) \\
\hline\hline 
\Delta  a_{\mu}(\sigma)    &~~303.3 ~(106.9)   \\
\hline
{{\rm lim}(95\%)} &93.8\le\delta a_{\mu} \le 512.8 \\      
\hline 
\end{array}   
$$

The   95\% CL interval $\delta a_{\mu}$ thus  obtained  
is positive, and hence
 it leads to  an {\sl allowed   positive} contribution 
({\sl an allowed  band}).
At the same time it also leads to  the {\sl exclusion} of any
 negative contribution  to the $a_{\mu}$. 
Note that the additional  positive contribution to $a_{\mu}$
can be up to a few times larger than the EW one.

Using   results for the vp1 contribution obtained  by the HMNT group \cite{tt}, 
from their 
``exclusive'' analysis, the values we get   for $a_{\mu}^{SM}$,
$\Delta  a_{\mu}(\sigma)$ and $\delta a_{\mu}$, 
are numerically very close to those obtained above, using 
the FJ02 analysis.
Keeping all the other contributions as before, but with the 
HMNT(ex) results for the vp1 term \cite{tt}, we get, in units of $10^{-11}$,
\beq
{\rm [HMNT (ex)]:}\hspace{1cm} \Delta  a_{\mu}(\sigma)= 297.0 \,(107.2) \hspace{0.5cm}
87.2 \le\delta a_{\mu} \le  507.4, 
\eeq
while  their   ``inclusive'' analysis  leads to a more stringent
constraint,  namely 
\beq
{\rm [HMNT (in)]:}\hspace{1cm} \Delta  a_{\mu}(\sigma)= 357.2 \,(106.4) \hspace{0.5cm}
148.7 \le\delta a_{\mu} \le  565.7.
\eeq

Note that
 the above 95\% CL intervals are  positive
 for both the HMNT results, just   as in  the FJ02 case.
 Also note  that  the relative differences between the 
upper bounds  in all  three analyses 
are small (below 10\%). However, the use of HMNT ``inclusive''
analysis  leads to a lower bound, which is  relatively 
 much higher (up to 70\%) than in the FJ02 and HMNT(ex) cases.

The other recently  published SM predictions \cite{Davier:1998si,
Narison:2001jt}, to which the BNL paper 
\cite{BNL2002} refers, were obtained  from  analyses
of the vp1 contribution based on the older, often not very precise 
$e^+e^-$ data. To improve the  accuracy of the estimation of the  vp1 part,  
  $\tau$ decay data were included in addition in those analyses. 
However,  
predictions obtained there for $a_{\mu}^{SM}$
and for $\sigma_{SM}$ are  not very different from these  
 new preliminary results used by us. 
It is worth mentioning that the following trend is 
observed (see discussion \eg in \cite{jeg02}): 
if  one uses the $\tau$-decay data in the calculation of vp1, 
then its estimation for the  value of vp1
 increases while that for  the  uncertainty decreases.
On the  other hand,  the preliminary FJ02 and HMNT analyses of the 
vp1 contribution, the ones used here, 
 rely solely on  the  low-energy data for the $e^+e^-$ collisions.
The accuracy of these analyses increases significantly with respect  to the 
earlier analyses of the same kind thanks  to 
an   inclusion of   new, high-precision measurements \cite{datalow} 
and the  use  of more refined theoretical methods.
Moreover,  the central value for the  vp1 contribution
did not increase in these new analyses. Thus, the    preliminary 
 analyses by FJ02 and HMNT lead   to a  deviation 
from the experimental value for the muon $(g-2)$ larger 
than the  published results mentioned above, which differ by ``only''
  1.6 to 2.6 $\sigma$, as pointed out in \cite{BNL2002}.

Let us conclude this general part by the following comment. It becomes
clear that already at present 
there is a  need of  a coherent and a comprehensive error analysis for all 
components contributing to the calculation of $\sigma_{\mu}^{SM}$.
It  will be even more nessesary  in a near future, 
when the BNL experiment will 
reach the planned accuracy. 
Since  an estimate combining all the different contributions does not exist
for $a_{\mu}^{SM}$, and hence for $\Delta a_{\mu}$,
which is  needed for estimating the effects of  new physics,
  is  not available at the moment,
a simplified error analysis, as  presented in this
section, is unavoidable.
This is  sufficient for a rough estimation of new effects;
however, to reach a final conclusion, better error analyses  are necessary.
\section{2HDM and existing constraints}
\subsection{The model}
The non-supersymmetric, CP-conserving  2HDM
(``Model II'') \cite{hunter}  
based on two doublets of complex, scalar-fields $\phi_1$,$\phi_2$. 
This is a simple extension of the SM,  in which  only the   Higgs sector 
is enlarged. To avoid possible  large effects due 
 to the flavour-changing neutral currents (FCNC),
the 2HDM potential can be chosen in  a  $Z_2$-symmetric form,
\ie without ($\phi_1$,$\phi_2$) mixing.
In the  general case, the potential  can have terms, characterized by a mass parameter $\mu$, which break  
the $Z_2$ symmetry softly; see \eg \cite{gin,san}.

The 2HDM   has  five Higgs particles: 
two neutral Higgs scalars $h$ and $H$, one neutral pseudoscalar $A$, 
and a pair of charged Higgses $H^{\pm}$.  Their masses
 are free parameters of the model. Other parameters are: 
the  angle $\alpha$, which describes the mixing in the neutral Higgs-scalar 
sector; $\tan \beta$ the ratio of  two vacuum expectation values
of scalar doublets, $\tan \beta= v_2/v_1$; and  the parameter $\mu$. 
Small values of $\mu$ seem to be more natural  from the   
point of view of the FCNC effects \cite{gin}. It is worth noticing that
for 
such a case,
a  non-decoupling of the  heavy Higgs sector can be realized
 \cite{gin,san}.
 
In the 2HDM one  can choose the Yukawa couplings in a few different ways. 
Here  we consider the  Model~(II) implementation, 
where one doublet of fundamental scalar fields couples to the $u$-type 
quarks, and the other to the $d$-type quarks and charged leptons. 
This way FCNC processes 
are avoided at the tree level \cite{weinberg,hunter}. This Higgs sector
is identical to the one in the MSSM; however,  in the  2HDM (II) considered by us,
there are no tree-level relations between parameters
as  in the MSSM case. Therefore even for very heavy supersymmetric particles,
the 2HDM (II) and MSSM  have very different phenomenology.

To be more specific, let us consider the ratios  
of the direct coupling constants of the Higgs boson
$h$ or $H$ to the massive gauge bosons $V=W$ or $Z$, as well as to the 
fermions 
{\it (\ie ~Yukawa couplings)} for the {$u$}-type quarks and {$ d$}-type
quarks and  the charged leptons, to the corresponding couplings  for the SM.
 They  are  determined in terms of  angles $\alpha$ and $\beta$
\cite{hunter,gin}. For $\chi_i^h \equiv 
{g_i^h}/{(g_i^h)_{SM}}$ (and similarly for $H$) we have,
in a form suitable for a   simultaneous discussion of $h$ and $H$:
\begin{eqnarray}
\chi_V^h=\sin(\beta-\alpha),\;\,\;& \chi_V^H=\cos(\beta-\alpha),& \chi_V^A=0,  \\
\chi_u^h=\chi_V^h+\cot\beta\chi_V^H,\;
&\chi_u^H=\chi_V^H-\cot\beta\chi_V^h,&\chi_u^A=-i \gamma_5\cot\beta, \\
\chi_d^h =\chi_V^h-\tan\beta\chi_V^H,\; &
\chi_d^H =\chi_V^H+\tan\beta\chi_V^h,\; & \chi_d^A=-i \gamma_5\tan\beta.
\label{2hdmcoup-h}
\end{eqnarray}
Here we have $(\chi_V^h)^2+(\chi_V^H)^2=1$.
Observe a {\sl pattern relation} between  these couplings
(for $h$ or $H$): $(\chi_u-\chi_V)(\chi_V-\chi_d)=1-\chi_V^2$, or 
$(\chi_u+\chi_d)\chi_V=1+\chi_u\chi_d$, found in  \cite{gin}.

For $\chi_V^h=1$ all couplings of $h$ have the SM values, 
the  couplings of $H$ to  gauge bosons are equal to zero, while 
one of the   couplings of $H$ to fermions may differ considerably 
from the corresponding SM one,  for a small or large $\tan \beta$. 
If $\chi_V^H=1$ then  the $H$-boson has  SM couplings, while 
$h$ has very different properties: $\chi_V^h=0$ and  the Yukawa coupling
 $\chi_d^h$ can be  large, for large values of $\tan \beta$.
This is a case that may correspond to a light-scalar scenario
discussed below.

The  Yukawa coupling  $\chi_d$, relevant to a 
Higgs boson coupling to a muon, plays a basic role in the calculation
of the 2HDM contribution to $a_{\mu}$.
It  is equal (up to a factor $-i\gamma_5$) to $\tan \beta$ for a pseudoscalar and $H^+$. 
If in addition $\chi_V^h=\sin (\beta-\alpha)=0$, then the same holds for a 
scalar $h$; 
more precisely then $|\chi_d^h|=\tan \beta$ (see eq. (7)).
In the calculation of the two-loop contribution to $a_{\mu}$,
the   coupling of $H^+$ to a scalar $h$ is involved as well, and it is given by  
\beq
\chi_{H^+}^h=\left(1-\frac{M_h^2}{2M_{H^+}^2}\right)\chi_V^h+\frac{M_h^2-\mu^2}
{2M_{H^+}^2}(\chi_d^h+\chi_u^h), 
\eeq 
with the same  normalization as that for an elementary charged scalar particle 
in the SM. For  $\chi_V^h=0$ one gets $\chi_{H^+}^h=(\chi_d^h -1/\chi_d^h)
 ({M_h^2-\mu^2})/({2M_{H^+}^2})$. We see that this coupling 
 depends on the parameter $\mu$. In this paper 
 we consider only the case with $\mu=0$.
A more general case will be studied elsewhere.
\subsection{Existing constraints}
Many searches for a light Higgs particle in the 2HDM (II) were performed
at various energies  and  machines; the most systematic 
studies were performed at LEP. 
All existing LEP data, see for instance   \cite{mk-lep,jembory,opal_mass,sin,yukawa}, 
allows for the  existence of   
{\sl one} light neutral Higgs boson, $h$ or $A$,  
with a mass even below 20 GeV.
According to the results presented in the left panel of Fig. 1, 
the other Higgs particle ($A$ or $h$, respectively)
should be heavy enough to avoid the exclusion region in the
($M_h,M_A$) plane, given roughly by $M_h+M_A\ge 90$ GeV.

This is  in contrast to the  case of the SM Higgs boson, which should 
be heavier  than 114.4
GeV (95\% CL). Also the MSSM Higgs particles should be heavier 
than $\sim$ 90 GeV \cite{jembory}.
An  analysis of the Bjorken process leads to an upper limit 
on the coupling of $h$ to the gauge boson $\chi_V^h$. This limit,
 obtained at  95\% CL, is presented
in the right panel of  Fig. 1. We see that  this coupling is much smaller than 1
for $M_h \lsim$ 50 GeV.
\begin{figure} 
\bce
\epsfig{file=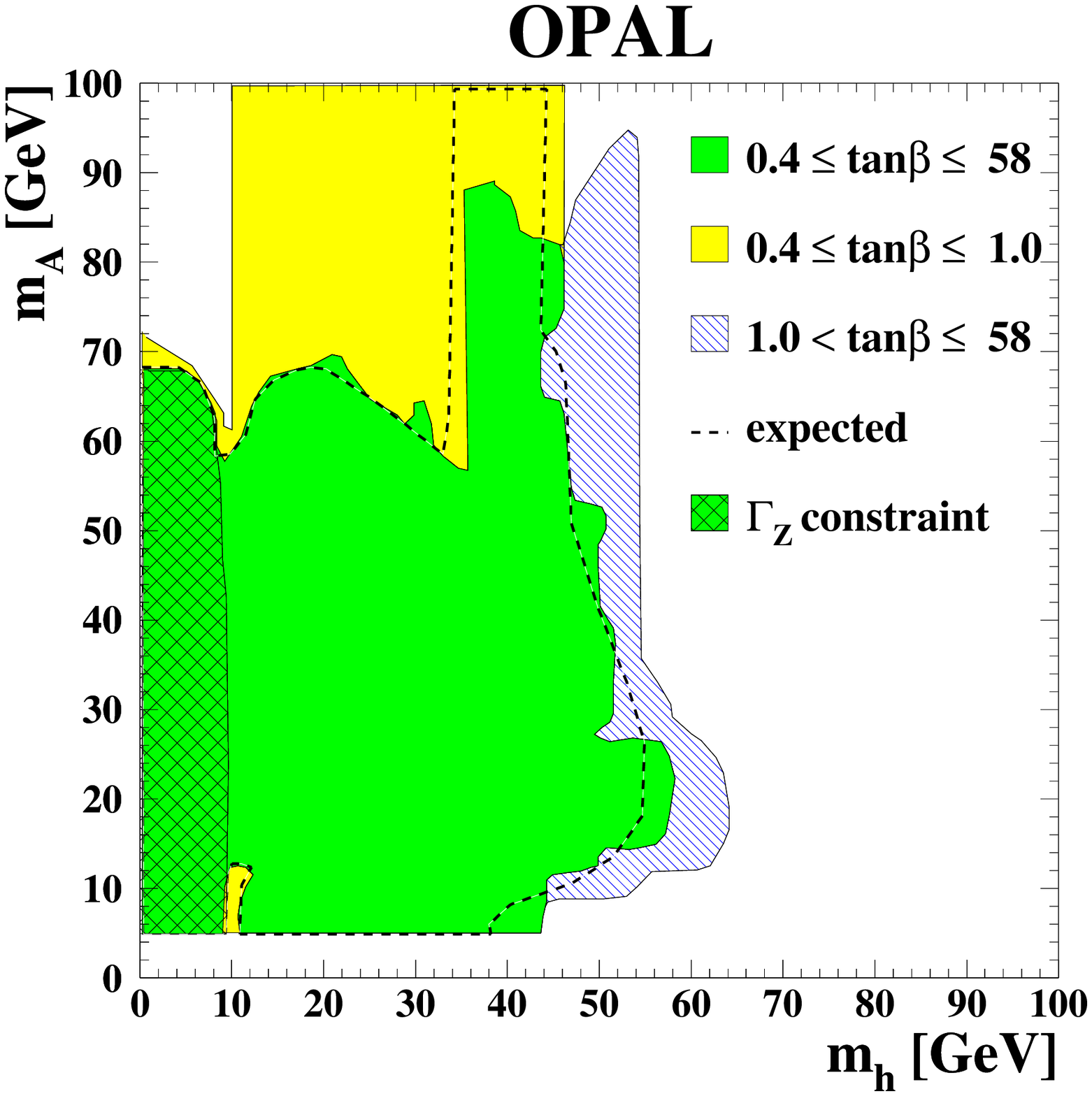,width=7.1cm}
\epsfig{file=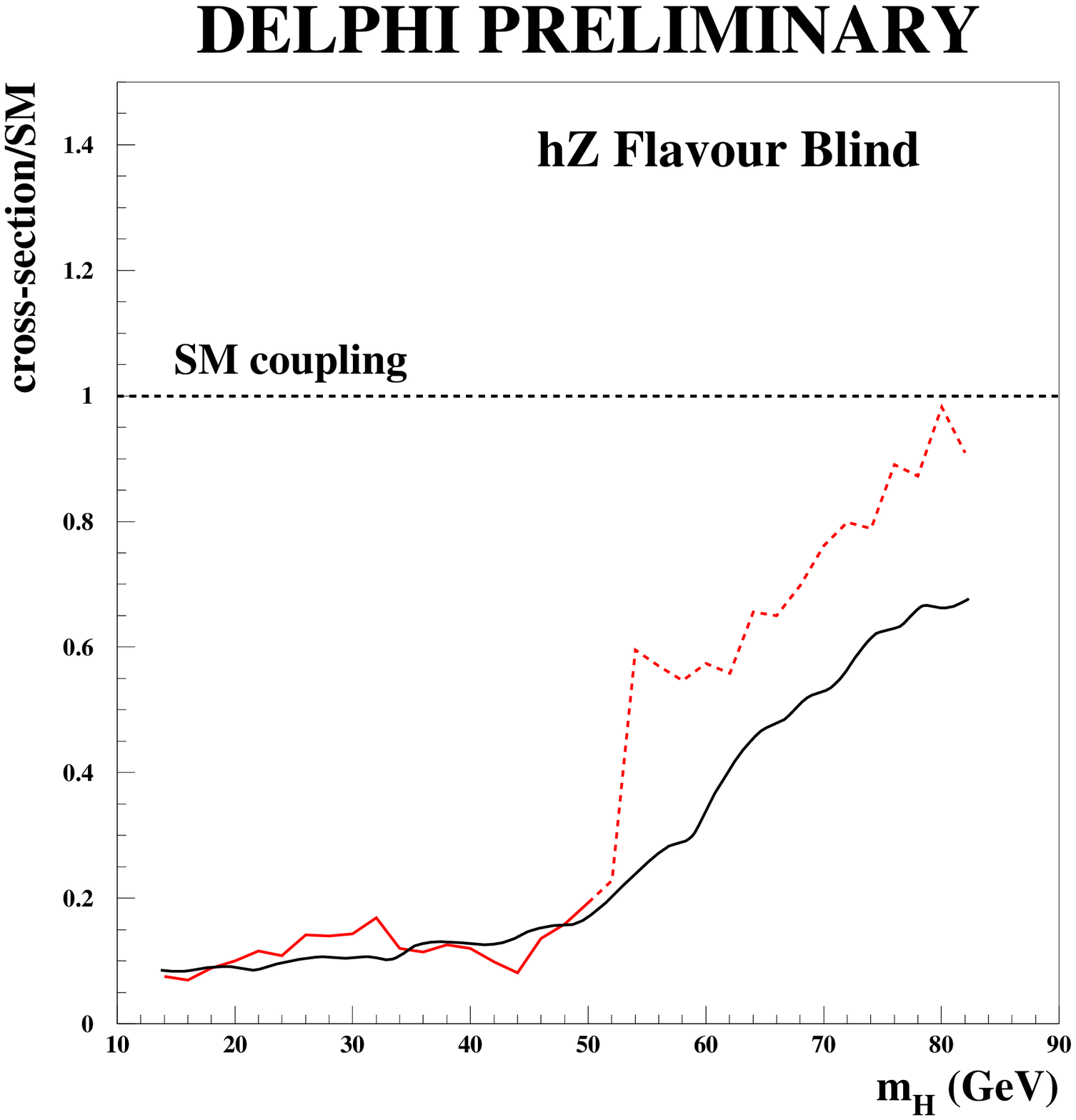, width=7.1cm}
\ece 
\caption{Left: The ($M_h,M_A$) exclusion plot from OPAL \cite{opal_mass}. 
Right: The upper limit on the  $(\chi_V^h)^2$  
(DELPHI preliminary results \cite{sin}).} 
\end{figure}
The Yukawa couplings $\chi_d$ of a  very light scalar or  of a  very 
light pseudoscalar,
with mass below 10 GeV,
are constrained in the form of upper limits 
  by the low-energy data \cite{keh,pich}, whereas 
 LEP experiments \cite{yukawa}  do that for for masses $\gsim$ 4 GeV 
(see Figs. 4, 5). 
It is only the analysis of the decay $Z\ra h/A \gamma$ at LEP \cite{mk-lep} 
that gives  both the upper and lower limits for $|\chi_d|$. Note
that $|\chi_d|$ is  
equal to  $\tan \beta$ for $A$ and, if $\chi_V^h=0$, also for $h$.

The  constraints from the $\Upsilon \ra h(A) \gamma$ process, mentioned above, 
have been measured  by a few groups \cite{keh}. 
We present their results  in  Fig. 4 (lines denoted by K, N and L). 
 Unfortunately, the corresponding predictions  have large experimental
and theoretical 
uncertainties, the latter ones
  both being due  to  the QCD and relativistic corrections.  
Nevertheless, as we will see below,
the constraints coming from this process,  even when   accounting for some
additional  uncertainties,  
 play an important role in   closing a low mass  
window for the  scalar $h$. 

Finally, note that in the 2HDM there  is an  important lower limit on the
 mass of $H^+$, coming from the NLO analysis of the $b \ra s \gamma$ 
data, given by   $M_{H^+}\ge 500$ GeV at 95\% CL \cite{Gambino:2001ew}.

\section{Constraining 2HDM(II) by $g-2$ for the muon data}

We apply the
 $\delta a_{\mu}$, obtained in sec. 2,
 to  constrain  parameters of the 2HDM (II) (see also earlier papers
\cite{mk-g2,Chang:2000ii,Dedes:2001nx}).
We  assume that the lightest Higgs boson, $h$ or  $A$, 
dominates the full 2HDM (II) contribution, $\ie$ we have 
$a_{\mu}^{2HDM} \approx a_{\mu}^h$, or $a_{\mu}^A$
 ({\it a simple approach} in \cite{mk-g2}). 
This approach should hold for  masses below 50 GeV, according to
 results presented in the left panel of Fig. 1. 
For  higher masses, which  are  also  considered here,
this is essentially equivalent  to   assuming  a large mass gap between 
the lightest one, $h$ or $A$, and the remaining Higgs bosons,
which lead to a light-$h$ or  a light-$A$ scenario. 
The relevant one- and two-loop diagrams, studied in 
\cite{old,mk-g2,Dedes:2001nx}
 and \cite{Chang:2000ii,mk-g2}, respectively,
are shown in Fig. 2 for the  $h$ and $A$ contributions. \\

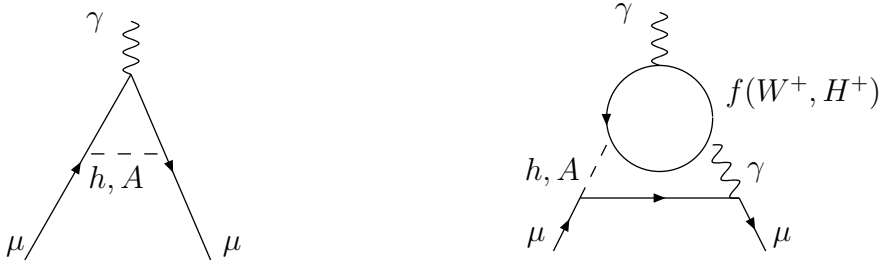
\begin{figure}[h]
\begin{center} 
\begin{picture}(300,80)(0,0)
\Text(30,90)[r]{$\gamma$}
\Text(45,30)[r]{$h,A$}
\Text(0,5)[r]{$\mu$}
\Text(82,5)[r]{$\mu$}
\Photon(40,90)(40,70){3}{3}
\ArrowLine(0,0)(40,70)
\ArrowLine(40,70)(70,0)
\DashLine(25,40)(50,40){5}
\end{picture} \\

\vspace*{0.2cm}
\begin{picture}(200,100)(-150,-110)
\Text(30,90)[r]{$\gamma$}
\Text(80,30)[r]{$\gamma$}
\Text(10,30)[r]{$h,A$}
\Text(-3,5)[r]{$\mu$}
\Text(90,5)[r]{$\mu$}
\Text(65,60)[l]{$f(W^+,H^+)$}
\Photon(40,90)(40,70){3}{3}
\Photon(60,40)(70,20){3}{3}
\ArrowLine(0,0)(10,20)
\ArrowLine(10,20)(70,20)
\ArrowLine(70,20)(80,0)
\ArrowArc(40,50)(20,2,1)
\DashLine(10,20)(20,40){5}
\end{picture} \\ 
\vskip -3.cm
\caption{ One- and two-loop ($W^+$ and  $H^+$
loops are only for a $h$-exchange) diagrams.}
\end{center}
\end{figure}

According to the LEP limits, discussed in sec. 3.2, 
we assume that $h$ does not couple  to $W/Z$, 
and therefore we neglect  the  $W$-loop  in the light-$h$ scenario.
We include, however, an $H^+$-loop  with $M_{H^+}$  equal to 400, 800 GeV 
 (and for  $\mu=0$).

In Fig. 3 we present the contributions to $a_{\mu}$  
obtained for an  $h$ (solid lines) and 
 for  a  $A$ (dashed lines),  assuming  Yukawa couplings 
$\chi_d$ equal to 1. 
For both $h$ and $A$, the  one-loop \cite{old,mk-g2,Dedes:2001nx} and two-loop 
\cite{Chang:2000ii,mk-g2} results
 are shown separately. For the purpose of comparison, 
 a one-loop $H^+$ contribution is presented. 
 The one-loop diagram gives a positive contribution to $a_{\mu}$ for  
a  scalar, whereas it is negative for
a  pseudoscalar, independently of the value of the Higgs-boson mass.
The signs of  the  two-loop contributions 
 are reversed, these diagrams contribute
negatively (positively) for  a $h$ ($A$) case \cite{Chang:2000ii}. 
 These two-loop diagrams  can give  large contributions,  since
they allow us to avoid one small Yukawa  coupling with a muon  in favor of 
a 
coupling with  other, potentially  heavy, particles circulating  
in the loop 
 \cite{Barr:1990vd,Chang:2000ii}. Indeed, the contributions of  
two-loop diagrams dominate over the corresponding one-loop ones
when the  mass of $h$ or $A$ is above a few GeV, as shown in Fig. 3.
As a result, in  the two-loop analysis, based on a sum of the one- and two-loop
 (fermionic and bosonic) contributions, 
a positive (negative) contribution can be ascribed
to a scalar $h$ with mass below (above) 5 GeV or  a  pseudoscalar $A$ with
mass above (below) 3 GeV. 

\begin{figure}
\bce
\epsfig{file=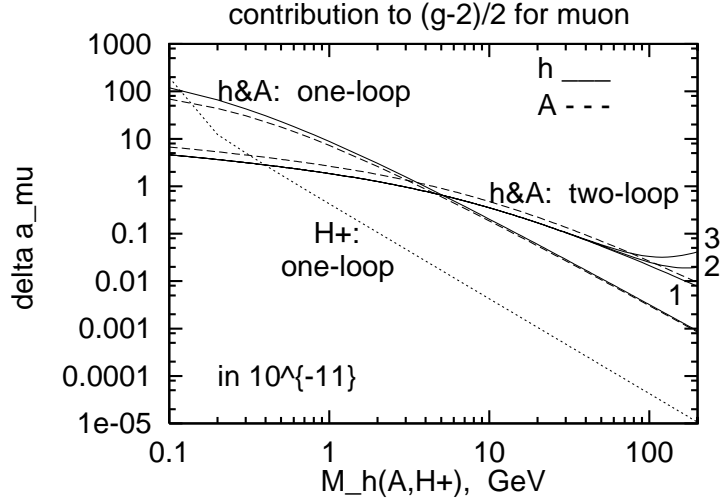, width=10cm}
\ece
\caption{ The (absolute value of)
individual contributions to $a_{\mu}$ from 
 a scalar $h$ (solid line),  a pseudoscalar $A$ 
(dashed line) and  a charged Higgs boson $H^+$ (dotted line). 
The one-loop  contribution for  $A$ and $H^+$, and  the two-loop one for $h$,  are  negative. Two-loop diagram contributions 
 for $A$ and $h$ (denoted ``1'') are
 based on the  down-type fermion loops only. 
For $h$ also, results with an additional contribution due to the
$H^+$-loop are shown:  line ``2'' (``3'') corresponds to $M_{H^+}$=800 (400) 
GeV.}
\end{figure}

In our calculation, $a_{\mu}^h$ and $a_{\mu}^A$ contain contributions 
proportional either  to $\chi_d^2$, equivalently to $\tan^2 \beta$ for $A$
and for $h$  (if $\chi_V^h=0$), or   to $\chi_d \chi_u=-1$.
Assuming $a^{h}_{\mu} = \delta a_{\mu}$ for a light-$h$ scenario,
and $a^{A}_{\mu} = \delta a_{\mu} $ for a light-$A$ one,
and using the estimate of the interval  $\delta a_{\mu}$  from table 2,
we can derive  constraints on $\tan \beta$, for $h$ and $A$.
They  are, as expected,  in the form of allowed regions (the 
area between thick lines in Fig. 4)
for masses below 5 GeV for $h$, shown in the left panel
of  Fig. 4,  and for masses above 3 GeV for $A$,
the right panel of Fig. 4 (see also \cite{Chang:2000ii}). 

\section{Combined 95\% CL constraints}

The 95\% CL constraints from the $(g-2)_{\mu}$
 are presented in Fig. 4 (area between thick  lines in the left and 
right panels) 
together with current upper limits     
from LEP, from the Yukawa processes \cite{yukawa} 
(see ALEPH, DELPHI and OPAL results). Also the 
 lower limits from the $Z\ra h(A) \gamma$ \cite{mk-lep} 
can be seen in Fig. 4. 
In addition, the upper 90\% CL limits  from 
 the  $\Upsilon$ decay (lines denoted K,N and L, with the 
 K results  rescaled by a factor 2, as discussed in
\cite{mk-g2}) and from the 
Tevatron \cite{TEV}, are presented as well.
\begin{figure}
\bce
\epsfig{file=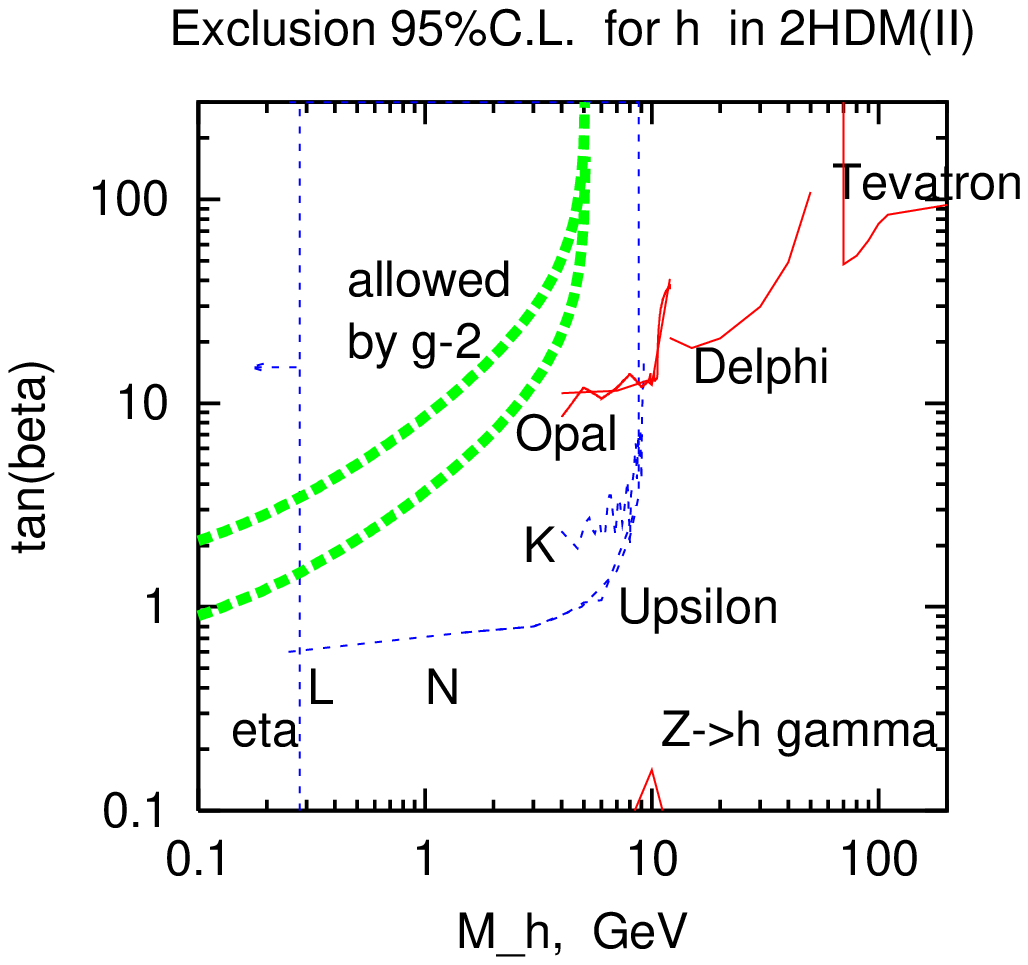, width=8.2cm}
\epsfig{file=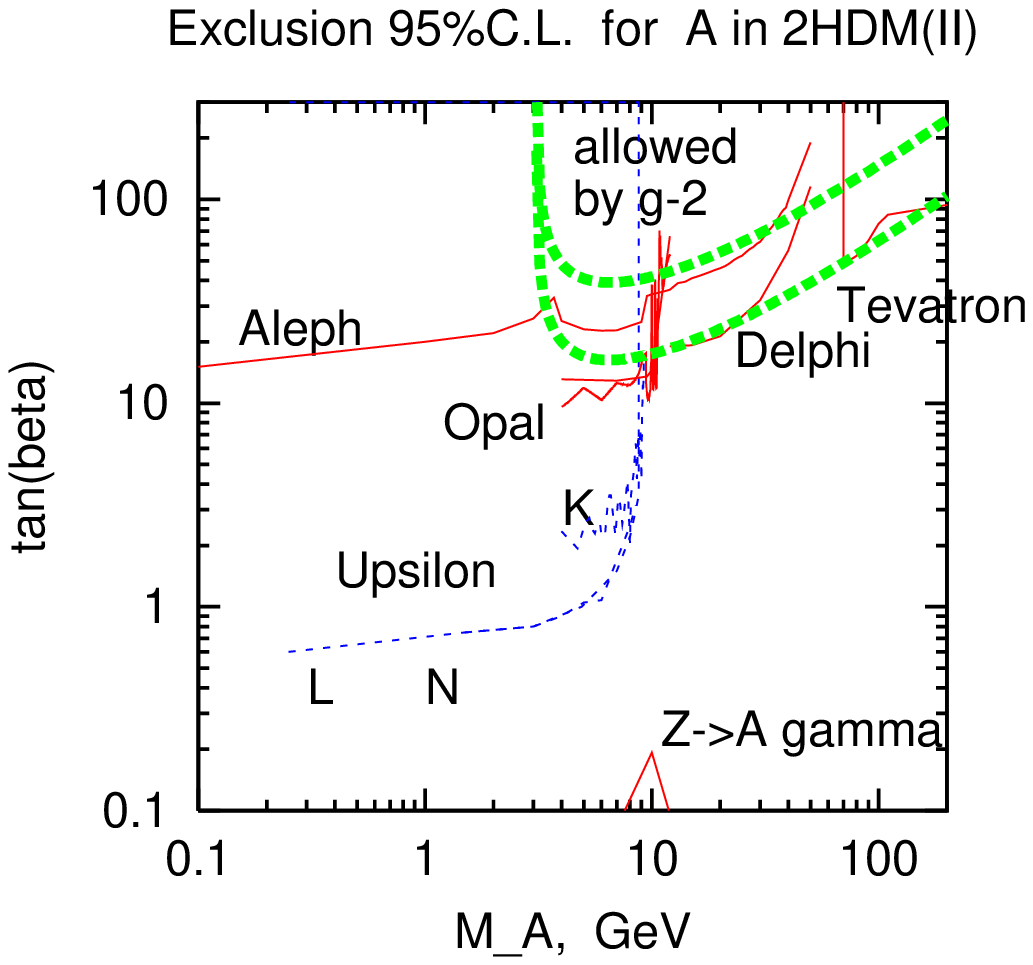, width=8.2cm}
\ece
\caption{Current 95\% CL constraints for $h$ (left) and for $A$ (right). 
The $(g-2)_{\mu}$ data (this analysis) give the
allowed area shown  between thick dashed lines. Also
the upper limits from LEP experiments (ALEPH, DELPHI,OPAL), Tevatron, 
as well as  from the $\Upsilon$ decay into $h(A) \gamma$ and the 
$\eta$ decay (for $h$ only) are presented. The 
 lower limits from LEP measurements on $Z\ra h(A) \gamma$ 
are shown as well. See text for details.} 
\end{figure}

We see that our two-loop analysis, based on the latest $(g-2)_{\mu}$ data 
and on the FJ02  estimation of  $a_{\mu}^{had}$ (vp1), 
 if combined with   constraints  from other experiments,
allows  for  the existence, in  the 2HDM (II), of a pseudoscalar 
with mass between $\sim$ 25 GeV and 70 GeV,
 and $\tan \beta$ above $\sim$ 25. The  mass
region for $A$ allowed by $(g-2)_{\mu}$ data, 
between $\sim$ 3 and  25 GeV, is  excluded by
the constraints from LEP, based on   OPAL and DELPHI data.
On the other hand the constraints from the  Tevatron  close 
the pseudoscalar-mass window above 70 GeV.
 For a  light scalar the combined constraints are even more severe; 
if the old constraints from the $\Upsilon$ decay data are taken into account,
the area allowed by the latest $(g-2)_{\mu}$ data  disappears. 
Note that the exclusion of a light $h$  is in agreement with the 
 conclusion of 
the  theoretical analysis reported in \cite{Kanemura:1999xf}.

\section{Conclusions and outlook}
Using the latest  precise measurement  of  the $(g-2)_{\mu}$   
and  comparing it
 with the improved,  theoretical estimations of the SM contribution,
we derive the 95\% CL $\delta a_{\mu}$ interval to be used to constrain
an additional contribution to $a_{\mu}$, beyond the SM ones. It
allows one to constrain strongly the  additional contribution, 
which arises in a CP-conserving, non-supersymmetric  2HDM (II)
for a small parameter $\mu$, studied in this paper. 
The additional contribution, allowed at 95\% CL, 
has to have a positive sign,
and can lead to a clear prediction for a light-scalar and a light-pseudoscalar
scenario in the  model considered here. These two scenarios correspond
to the case when one of the Higgs  bosons, $h$ or $A$, is very light, 
 much lighter than the other Higgs particles of the model.
It should be further noted that both of these scenarios are 
in agreement with existing data from other experiments.
An exchange of such light particle dominates in the 
one- and two-loop contributions to the $a_{\mu}$.
Constraints from $(g-2)_{\mu}$ are such that they exclude  
a light $h$  with a mass above 5 GeV and a light $A$
if its  mass is below 3 GeV, as
the corresponding contributions are negative in these regions.

Our  two-loop analysis presented in this paper
is based on the newest $(g-2)_{\mu}$ data 
and on the  (preliminary) FJ02  estimation of  $a_{\mu}^{had}$. 
Combining  the constraints from the $(g-2)_{\mu}$ data with those from 
 other experiments,    
a pseudoscalar with mass between $\sim$ 25 ~GeV and 70 ~GeV
 and $25 \lsim \tan \beta \lsim 115$ is allowed. However, a light scalar 
is excluded.

The main results will hold also if the $g-2$ constraints are  based on 
the HMNT results for the vp1. For the HMNT(in) case 
the window for a pseudoscalar will be even smaller:
$35 \le M_A \le 70$ GeV and  $\tan \beta$ between 40 and 120.    
The results will not change significantly, if the uncertainty
for the lbl contribution is up  two times larger than in 
the estimation we used in the analysis.
If such an error is still added in quadrature, then the lower bounds go down, 
with  respect to  results given in Fig. 4, by a factor $\sim$ 1.4. There will be 
no visible changes in the upper bounds.

Finally, we stress  a need for   a coherent and  comprehensive error 
analysis for the SM contributions to the $a_{\mu}$ (see a very recent paper
[44]).   

\vskip 0.3cm
\noindent
{\large \bf {Acknowledgements}}\\
I am grateful to Fred Jegerlehner and Thomas Teubner
for many valuable discussions  of the recent results on 
the $g-2$ for the muon. I am especially indebted to Rohini Godbole 
for critical comments  and important suggestions.   I thank also 
 S\l awek Tkaczyk and other colleagues
for useful  discussions and information.
I am grateful to M. Boonekamp, M. Kobel and F. Akesson 
for sending  me  results on the Yukawa process and on other searches at LEP.  
I acknowledge the important contribution by Suzy Vascotto.
Finally, I would like to thank the organizers of this excellent 
and a very special meeting for their help.

\end{document}